\documentclass[aps, prl, superscriptaddress, reprint, nobibnotes, twocolumn]{revtex4-2}
\usepackage{amsmath,amsthm,amssymb,mathtools,physics} 
\usepackage{xcolor} 
\usepackage{graphicx} 
\usepackage[font=small, justification=raggedright, singlelinecheck=false]{caption}
\usepackage{subcaption} 
\usepackage{bm} 
\usepackage{lineno} 
\usepackage{geometry} 
\usepackage{adjustbox} 
\usepackage{placeins} 
\usepackage[T1]{fontenc} 
\usepackage{lipsum} 
\usepackage{csquotes} 
\usepackage{enumerate} 
\usepackage{epstopdf} 
\usepackage{fancyhdr} 
\usepackage{marvosym} 
\usepackage{wrapfig} 
\usepackage{indentfirst} 
\usepackage{balance} 

\def\bef{\begin{framed}}
\def\eef{\end{framed}}
\def\be{\begin{equation}}
\def\ee{\end{equation}}
\def\ber{\begin{eqnarray}}
\def\eer{\end{eqnarray}}

\def\rv{{\bf r}}

\def\vv{{\bf v}}
\def\mv{{\bf m}}

\def\kv{{\bf k}}

\def\fv{{\bf f}}

\def\sigmav{\bm{\sigma}}

\usepackage{lineno}

\usepackage[pdftex, pdftitle={Article}, pdfauthor={Author}]{hyperref}
\hypersetup{
  breaklinks = true,
  citecolor = blue,
  colorlinks = true,
  linkcolor = red,
}

\begin{document}
\title{Insulator-Metal Transition and Magnetic Crossover in Bilayer Graphene}

\author{Amarnath Chakraborty}
\email{achakraborty@mail.missouri.edu}
\affiliation{Department of Physics and Astronomy, University of Missouri, Columbia, Missouri, USA}

\author{Aleksandr Rodin}
\email{aleksandr.rodin@yale-nus.edu.sg}
\affiliation{Yale-NUS College, 16 College Avenue West, 138527, Singapore}
\affiliation{Centre for Advanced 2D Materials, National University of Singapore, 117546, Singapore}
\affiliation{Department of Materials Science and Engineering, National University of Singapore, 9 Engineering Drive 1, 117575, Singapore}

\author{Shaffique Adam}
\affiliation{Department of Physics, Washington University in St. Louis, St. Louis, Missouri 63130, United States}
\affiliation{Department of Materials Science and Engineering, National University of Singapore, 9 Engineering Drive 1, 117575, Singapore}

\author{Giovanni Vignale}
\email{vgnl.g@nus.edu.sg}
\affiliation{The Institute for Functional Intelligent Materials (I-FIM), National University of Singapore, 4 Science Drive 2, Singapore 117544}
\date{\today} 

\begin{abstract}
In-plane magnetic fields offer a relatively unexplored opportunity to alter the band structure of stacks of 2D materials so that they exhibit the desired physical properties. Here we show that an in-plane magnetic field combined with a transverse electric field can induce an insulator-metal (IM) transition in bilayer graphene. Our study of the magnetic response reveals that the orbital magnetic susceptibility changes from diamagnetic to paramagnetic around the transition point. We discuss several strategies to observe the IM transition, switch the diamagnetism, and more generally control the band structure of stacked 2D materials at experimentally accessible magnetic fields.

\end{abstract}

\maketitle

\textit{Introduction.}
The effects of perpendicular magnetic fields on the electronic properties of monolayer graphene have been studied in a variety of contexts over the years.  Landau levels\cite{Zhang2005}, integer \cite{Novoselov2006} and fractional\cite{Bolotin2009} quantum Hall effects were demonstrated experimentally within several years of graphene discovery\cite{Novoselov2005}. Very recently, a giant magnetoresistance of Dirac electrons in high-mobility graphene has been discovered \cite{Xin2023}. 
In addition, it has been shown that the orbital magnetic response of graphene can be controlled by modifying the number of charge carriers via gating.  Perfect diamagnetism at the charge neutrality point \cite{PhysRev.104.666,PhysRev.119.606,PhysRevB.9.2467} is superseded by paramagnetism as the Fermi level approaches a van Hove singularity in the density of states \cite{PhysRevLett.67.358} making it possible to switch the monolayer between diamagnetic and paramagnetic regimes\cite{PhysRevLett.112.026402}. All of these effects have been experimentally observed \cite{doi:10.1126/science.abf9396} and thoroughly investigated theoretically  \cite{Koshino11,PhysRevLett.100.236405,PhysRevLett.104.225503}.

In recent years, bilayer and multilayer graphene have attracted significant interest due to their highly tunable properties under mechanical strain, twisting, and external fields. Their remarkable orbital magnetic properties \cite{PhysRevB.76.085425, Koshino11} and the discovery of flat bands in twisted bilayer graphene, which host superconductivity \cite{Cao2018}, further highlight their importance. A recent study on Bernal-stacked bilayer graphene reveals a complex multi-cone band structure under finely tuned displacement fields \cite{Seiler2024}. Another important property of this system is its strong response to in-plane magnetic fields, which couple to the in-plane component of the orbital magnetic moment through the interlayer looping motion of electrons \cite{Kammermeier_2019, PhysRevB.89.125418, PhysRevB.88.241107, PhysRevB.93.115423,Park2022}. In a bilayer system, for instance, the vector potential associated with the in-plane magnetic field can have opposite signs in the two layers, while remaining constant in each layer. This causes the band structures of the layers (before turning on interlayer coupling) to shift in opposite directions in momentum space, leading to non-trivial modifications of the band structure when interlayer coupling is considered. This effect was experimentally demonstrated a few years ago in twisted bilayer graphene, where an in-plane magnetic field was used to alter the separation of nearby Dirac cones\cite{doi:10.1126/science.aaf4621}. 

However, despite this pioneering work, the potential of in-plane magnetic fields as a tool for band-structure manipulation remains largely unexplored. In particular, no attention has been paid to the possibilities opened by the interplay between an in-plane magnetic field and an electric field perpendicular to the layer, the latter being commonly used as a tool to change the band gap~\cite{PhysRevLett.99.216802,Zhang2009}.  

This paper aims to describe some of these possibilities. We focus, for clarity, on the simplest model of a non-twisted, Bernal-stacked graphene bilayer \cite{RevModPhys.81.109, McCann_2013, PhysRevLett.115.015502, Yan2011} and show that it can be driven through a transition from a gapped insulator to a compensated semimetal by the application of a very strong in-plane magnetic field combined with a vertical displacement field. This phenomenon is just an example of the kind of control that can be exerted on the properties of bilayer graphene by means of crossed magnetic and electric fields: the magnetic field controls the $k$-space separation of the Dirac cones arising from each layer, while the electric field controls their displacement along the energy axis.
Second, we show that an even richer scenario emerges when we consider the orbital magnetic response to an in-plane magnetic field as a function of doping and electric bias. 
We find that the in-plane orbital magnetic susceptibility (OMS) switches from diamagnetic to paramagnetic concomitantly with the field-induced IM transition at half-filling. Moving away from half-filling we find that the in-plane magnetic response can be switched from diamagnetic to paramagnetic by adjusting the displacement field and the position of the chemical potential within the bands. We also demonstrate a clear correlation between the amplitude of the fluctuations of the in-plane magnetic moment and the magnitude of the paramagnetic response \cite{SI}.


Taken together, our findings reveal an unexpected and hitherto unnoticed interplay between electronic transport and in-plane magnetism in layered electronic systems.  Crucially, this interplay can be controlled by modest displacement fields on the order of 100 meV, 
even though the value of the in-plane magnetic field required to achieve sizeable effects -- while formally proportional to the interlayer coupling -- is unrealistically high ($\sim$ 1000 T for bilayer graphene). 
To remedy this, we propose several strategies for applying the controlling fields to artificially strained and/or twisted multilayer structures. These structures are indeed considered most promising for band structure engineering, which targets desired physical properties.  
With Brillouin zones up to 100 times smaller than the ``natural" ones, these artificial structures should exhibit the predicted effects at much smaller values of the in-plane field --  as low as 10 T.

\textit{Model.}
The Bernal-stacked graphene bilayer is described by a tight-binding Hamiltonian with nearest-neighbor intra- and interlayer hopping~\cite{RevModPhys.81.109}
\begin{align}
H_{\mathbf{k}}& =\begin{pmatrix}
     V & \gamma_{0} f\left(\mathbf{k}_\mathrm{t}\right) & 0 & 0\\
    \gamma_{0} f^{\dagger}\left(\mathbf{k}_\mathrm{t}\right) & V &\gamma_{1} &0\\
    0 & \gamma_{1} & -V & \gamma_{0} f\left(\mathbf{k}_\mathrm{b}\right) \\
   0 & 0 &\gamma_{0} f^{\dagger}\left(\mathbf{k}_\mathrm{b}\right) &-V
\end{pmatrix}
\label{eqn:Hamiltonian}
\end{align}
The four basis states are $\mathrm{A}_\mathrm{t}$, $\mathrm{B}_\mathrm{t}$, $\mathrm{A}_\mathrm{b}$, and $\mathrm{B}_\mathrm{b}$ with $\mathrm{A}$ and $\mathrm{B}$ denoting the sublattice and $\mathrm{t}/\mathrm{b}$ subscripts identifying the top and bottom layers.
The hopping energy between $\mathrm{B}_\mathrm{t}$ and $\mathrm{A}_\mathrm{b}$ is $\gamma_1 = 0.3$eV, while the hopping term between the nearest neighbors within a layer is $\gamma_0 = -2.8$eV \cite{McCann_2013}.
$f(\mathbf{k}) = \sum_{n = 1}^3e^{i\mathbf{k}\cdot\boldsymbol{\delta}_n}$, where $\boldsymbol{\delta}_{1} = a(-1, \sqrt{3})/2$, $\boldsymbol{\delta}_{2} = a(-1, -\sqrt{3})/2$, and $\boldsymbol{\delta}_{3} = a(1,0)$ are the vectors connecting atoms $\mathrm{A}$ to their nearest neighbors within a layer and $a\approx1.42$\AA\ is the carbon bond length in graphene.
The diagonal terms $\pm V$ originate from an external electric field applied perpendicular to the bilayer (the displacement field).
Finally, the constant vector potential is incorporated into the Hamiltonian by using the Peierls substitution $\mathbf{k}\rightarrow \mathbf{k} \pm e\mathbf{A} / \hbar$, so that we have
\begin{equation}
   \mathbf{k}_\mathrm{t/b} a
    =  \mathbf{k} a \pm \pi\Phi (\sin\theta,-\cos\theta,0)\,,
    \label{eqn:Peierls}
\end{equation}
where $l$ is the interlayer separation and $\Phi = B al/\Phi_0$ is the magnetic flux through a rectangle defined by an intra-plane bond and an inter-plane segment in units of the magnetic flux quantum  $\Phi_0 = h/e\simeq 4.14 \times 10^5$ T.\AA$^2$.
Here $\theta=\pi/2$ corresponds to a vector potential parallel to the $x$-axis (zig-zag direction) and a magnetic field parallel to the $y$-axis (armchair direction). 
We can easily show (see SM \cite{SI}) that the eigenvalues of $\hat H_{\kv}$ change sign under reversal of $\kv$ -- other parameters remain unchanged. This implies that the system at half-filling can either be an insulator -- if there are no zero eigenvalues -- or a compensated semimetal with equal densities of electrons and holes in pockets below and above the Fermi level. 
\begin{figure}
    \centering
    \includegraphics[width=\columnwidth]{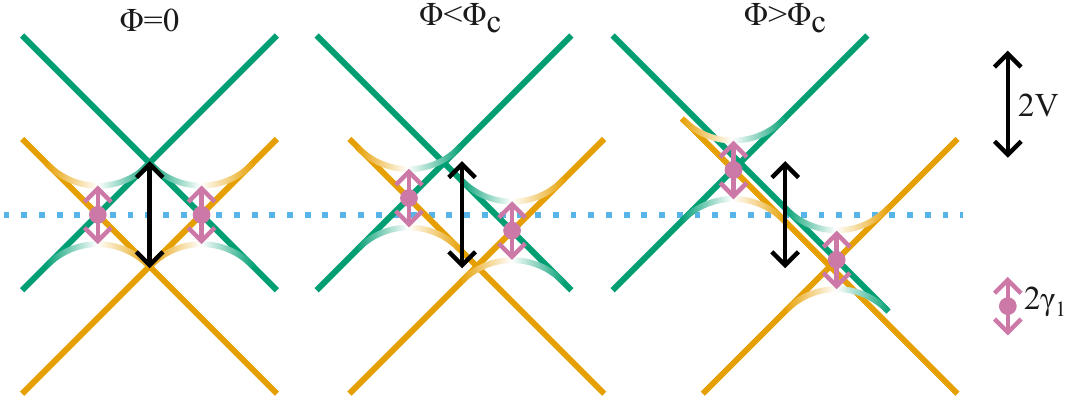}
    \\
    \includegraphics[width = \columnwidth]{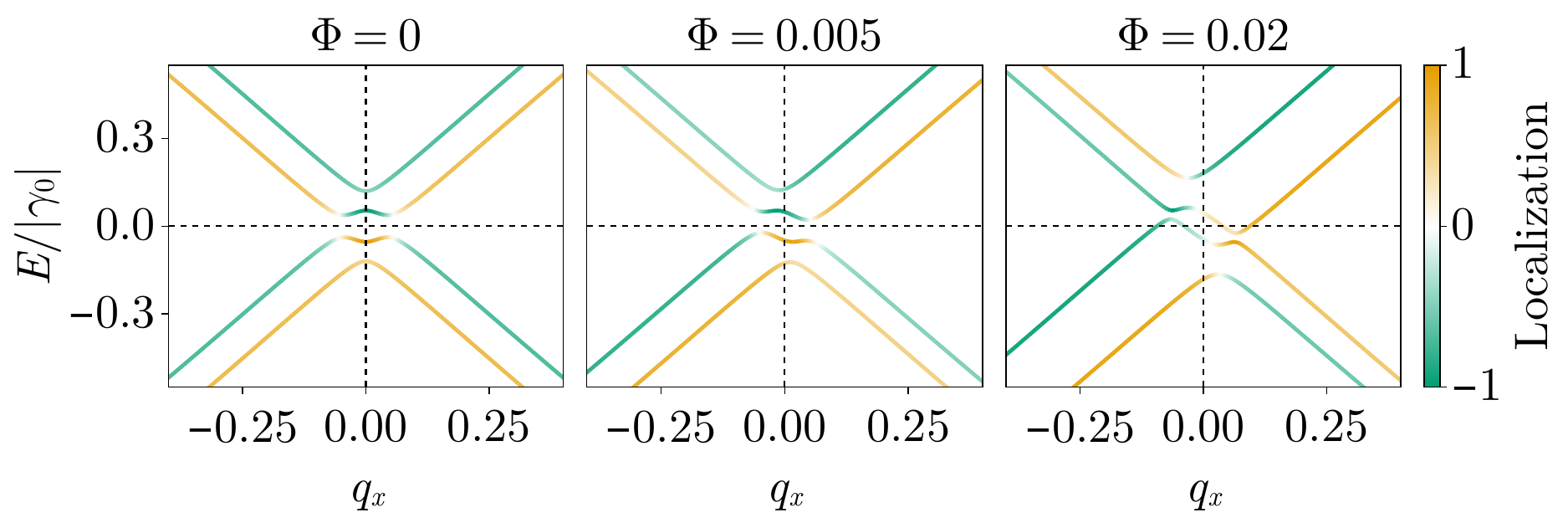}
    \caption{Top row: Schematic illustration of the evolution of the bilayer band structure as a function of $\Phi$ at finite $V$.
    The two colors correspond to the unperturbed Dirac cones from the two layers. The dashed lines show the level of repulsion due to inter-layer coupling. The IM transition occurs when the upper anticrossing level on the right and lower one on the left move respectively below and above the chemical potential at half-filling.
    Bottom row: Numerically computed bands close to the K point $\left(2\pi / 3, 2\pi/\sqrt{3} \right)$ with $V = \gamma_1 / 2$ demonstrating the insulator-metal transition at $\Phi_c = |\gamma_1/3\pi\gamma_0| \approx 0.011$.
    The direction of the field for all plots is $\theta = \pi / 2$.
    The color of the bands shows the distribution of the states over the layers, as defined in the text.}
    \label{fig:Bands}
\end{figure}
\textit{Insulator-metal transition.}
We now show that an IM transition (or, more precisely, a transition from insulator to compensated semimetal) must necessarily occur at a critical value of $\Phi$ at half-filling.  The top panel of Fig.~\ref{fig:Bands} shows schematically the band structures of the two non-interacting monolayers (green and orange cones for top and bottom, respectively) in the vicinity of a Dirac point.
The two cones are displaced vertically along the energy axis by the potential $V$ and horizontally along the momentum axis by the in-plane field $\Phi$.
Importantly, the crossing between levels belonging to different layers moves above and below the chemical potential by an amount $2\pi v\Phi$ where $v=3|\gamma_0|/2$ is the velocity of the Dirac fermions with lengths given in terms of the carbon bond length $a$.
With the interlayer coupling switched on, one of the crossing levels is pushed up by $\gamma_1$ while the same amount pushes down the other.
The system remains an insulator {\bf (possibly a zero-gap insulator in the special case $V=0$)} as long as $\gamma_1<3\pi|\gamma_0|\Phi$ and switches to compensated semimetal when this inequality is violated.
This simple argument allows us to pinpoint the transition at
\begin{equation}
    \Phi_c=\frac{\gamma_1}{3\pi|\gamma_0|}
\label{eqn:crit_field}
\end{equation}
in perfect agreement with more accurate analytical treatment, as shown in SM \cite{SI}. Notice that $\Phi_c$ does not depend on the direction of the in-plane field. 

In addition to providing a schematic illustration, we calculate the bands close to the K point and show how they change with $\Phi$. \footnote{
All computations in this paper are performed using {\scshape julia}~\citep{Bezanson2017}.
The plots are made with Makie.jl package~\citep{Danisch2021} using the color scheme designed for colorblind readers~\citep{Wong2011}.
The scripts used for computing and plotting can be found at \href{https://github.com/rodin-physics/bilayer-graphene/tree/main}{Bilayer graphene in crossed field setup}}
Comparing the results at the bottom of Fig.~\ref{fig:Bands} to the schematics in the top row, we observe a good qualitative agreement between the two.
For the parameter values given following Eq.~\eqref{eqn:Hamiltonian}, $\Phi_c \approx 0.011$.
Indeed one can see that, for $\Phi < \Phi_c$, the numerical calculations result in a gapped bandstructure, while $\Phi > \Phi_c$ yields a compensated semimetal.

We pause here to comment on the peculiar nature of the semi-metallic state emerging from the IM transition.  Due to the simultaneous presence of two types of carriers at the Fermi level its transport properties are expected to be strongly affected by electron-hole scattering resulting in reduced electrical conductivity and enhanced thermal conductivity \cite{PhysRevB.102.214304,cruise2024observability}. The system is also a good candidate for observing the elusive two-fluid hydrodynamics \cite{PhysRevB.97.121404,PhysRevB.101.045421}.

Although the critical magnetic field is formally proportional to the interlayer coupling $\gamma_1$ -- and thus can be exponentially reduced by increasing the spacing between the layers -- its numerical value is still very large. With commonly accepted values of $a$ and $l$ we find $B_c = |\gamma_1 /3\pi\gamma_0|\Phi_0 / al \approx 1000$~T. In the concluding section, we will discuss several effects that could bring this large value down to a more accessible range.

\textit{Magnetic Response.} Next, we address the magnetic susceptibility 
\begin{equation}
 \chi = \frac{\mu_{0}}{\mathcal{V}} \frac{d^2\Omega}{dB^2}\,,
     \label{eqn:Mag_Quantities}
\end{equation}
where $\mu_0$ is the magnetic constant, $\mathcal{V}$ is the area of the system, and $\Omega$ is the Helmholtz free energy given by 
\begin{align}
    \Omega(\Phi,\theta,V,\mu,T) 
    = -T\sum_{\mathbf{k},n} \ln\left[1 + e^{-\frac{E_{n}(\mathbf{k}, \Phi,\theta,V)-\mu}{T}}\right]\,.
    \label{eqn:Omega}
\end{align}
Here, $E_{n}(\mathbf{k}, \Phi,\theta,V)$ is the energy of the $n$th eigenstate at wave vector $\kv$, temperature $T$, and chemical potential $\mu$.
To compute $\chi$, we first set $\Phi \rightarrow \Phi + \delta$ and Taylor-expand the Hamiltonian to the second order in $\delta$ which yields $H_\mathbf{k} \simeq H_{0,\mathbf{k}}+\delta H_{0,\mathbf{k}}' + \frac{1}{2}\delta^2 H_{0,\mathbf{k}}''$, where $H_{0,\mathbf{k}}$ is the Hamiltonian at $\delta = 0$, $H_{0,\mathbf{k}}'$ and $H_{0,\mathbf{k}}''$ are its first and second derivatives also evaluated at $\delta=0$.
Using perturbation theory, we expand the eigenvalues $E_n$ to second order in $\delta$, insert them in Eq.~\eqref{eqn:Omega} and expand again to second order in $\delta$ to extract the susceptibility:
\begin{align}
    \chi 
   &= -\frac{\mu_0}{\mathcal{V}}\left(\frac{al}{\Phi_0}\right)^2
    \sum_{\mathbf{k}, n}
    \Bigg[ 
    n_F(E_{n}^0(\mathbf{k}))
    \langle \mathbf{k}, n|H''_{0,\mathbf{k}}|\mathbf{k}, n\rangle
    \nonumber
    \\
    &
    +
    \sum_{m\neq n}
    \frac{n_F(E_{n}^0(\mathbf{k}))-n_F(E_{m}^0(\mathbf{k}))}{E_{n}^0(\mathbf{k}) - E_{m}^0(\mathbf{k})}
    |\langle \mathbf{k}, n|H'_{0,\mathbf{k}}|\mathbf{k}, m\rangle|^2
    \Bigg]
\nonumber
\\
     &
     - \frac{\mu_0}{\mathcal{V}}\left(\frac{al}{\Phi_0}\right)^2\sum_{\kv,n}n_F'(E_{n}^0(\mathbf{k}))
   |\langle \mathbf{k}, n|H'_{0,\mathbf{k}}|\mathbf{k}, n\rangle|^2 \,
   \label{eqn:Chi}
\end{align}
where $|\mathbf{k},n\rangle$ are eigenvectors of $H_{0,\mathbf{k}}$, $E^0_{n}(\mathbf{k})$ are the corresponding eigenvalues,  $n_F$ is the Fermi distribution function and $n_F'$ its derivative with respect to energy.
The sum over momenta in the first Brillouin zone is converted into an integral $\mathcal{V}^{-1}\sum_\mathbf{k} \rightarrow(2\pi a)^{-2} \int_0^{\frac{4\pi}{3}} dk_x \int_0^{\frac{2\pi}{\sqrt{3}}} dk_y $. 
\begin{figure}
    \centering
    \includegraphics[width =\columnwidth]{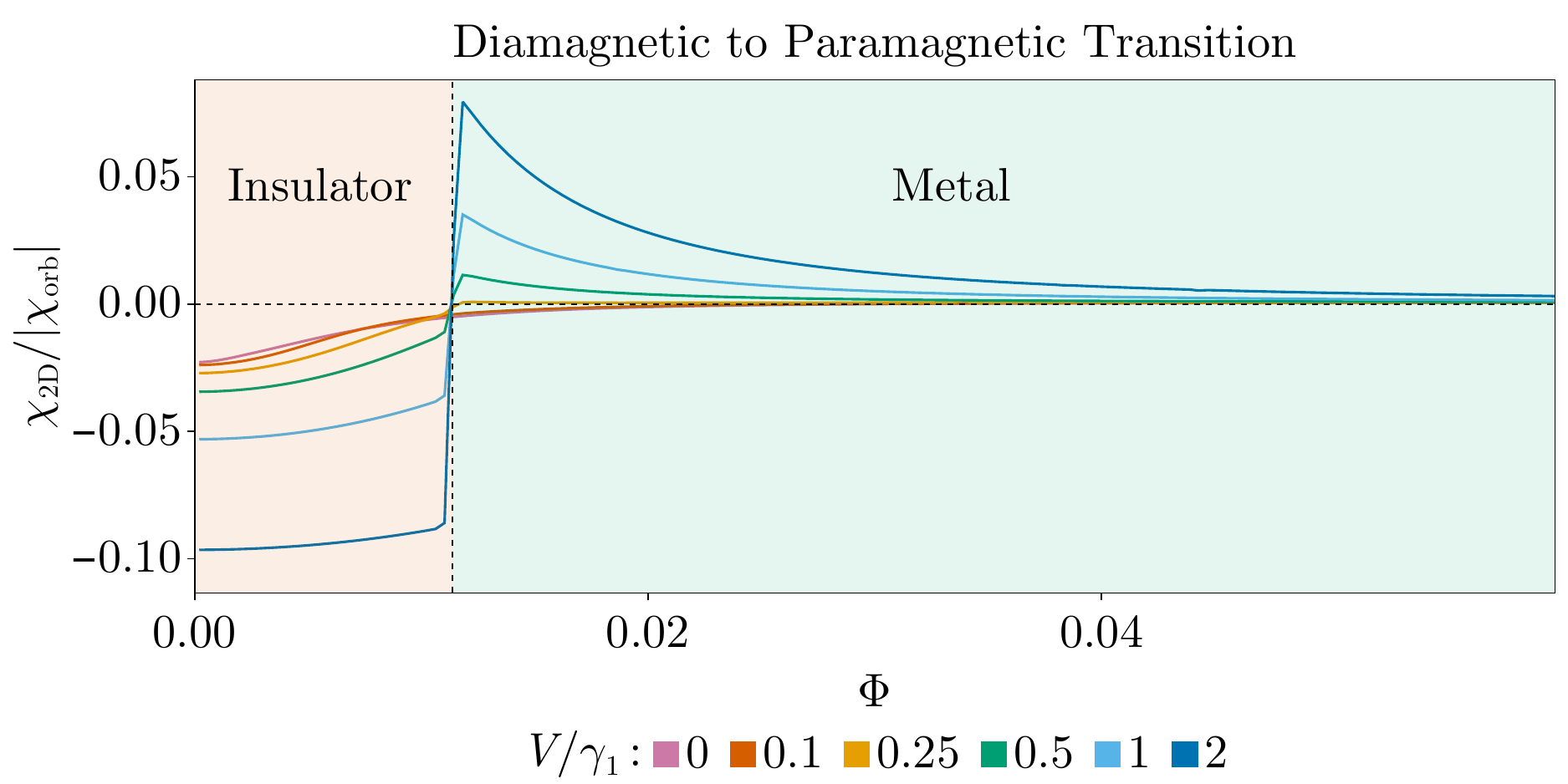}
\caption{Plot of the orbital magnetic susceptibility (OMS) vs in-plane magnetic flux \(\Phi\) at $\mu=0$. The calculation is done at \(T=12\)K. 
Initially diamagnetic in the insulating phase, the susceptibility rapidly becomes paramagnetic at the critical field. 
The size of the jump in the OMS increases with increasing voltage $V$ between the layers and vanishes for $V \to 0$.  
} 
\label{fig:Chi_vs_Phi}
\end{figure}
Notice that the 2D susceptibility has the dimensions of a length.  To give a clearer idea of its magnitude we divide the numerically calculated $\chi_{2D}$  by the absolute value of the out-of-plane orbital susceptibility of a charge-neutral  graphene monolayer at $T=300$K \cite{PhysRev.119.606,PhysRev.104.666,PhysRevB.9.2467} i.e.,   
$|\chi_{orb}|=\frac{1}{6\pi} \frac{e^2v^2}{c^2} \frac{1}{k_B T}\approx 2.7 \times 10^{-14} m$.
 The ratio $\frac{\chi_{2D}}{|\chi_{orb}|}$ is plotted as a function of $\Phi$ at $\mu = 0$ in Fig.~\ref{fig:Chi_vs_Phi} for several values of $V$.
The most striking feature exhibited by the plots is the sharp transition from diamagnetic to paramagnetic behavior at the point of IM transition.
It is possible to understand the origins of this behavior by considering individual terms in Eq.~\eqref{eqn:Chi} for $T \rightarrow 0$.
The first term is negative, i.e., diamagnetic, and has a very weak dependence over the chemical potential. The second and third terms are both positive and therefore paramagnetic.
Specifically, the second term describes inter-band paramagnetism~\cite{Ashcroft76} and the final term is a Fermi contour contribution that arises from the direct coupling of the in-plane magnetic field with the fluctuating orbital moment (\cite{SI}): this term vanishes when the system is an insulator.

Clearly, the diamagnetic contribution dominates in the insulating phase. The sharp switch to paramagnetism that accompanies the IM transition results from the irruption of the Fermi surface term.  The subsequent decrease of $\chi$ with a further increase in $\Phi$ suggests that the magnitude of the paramagnetic response strongly depends on the nature of the contributing states at the Fermi level. 
A closer look at the bottom row of Fig.~\ref{fig:Bands} with $\Phi = 0$ reveals that when the Fermi level is close to the band edge the contributing states are essentially {\it delocalized}, i.e., evenly distributed over the two layers.
This implies large inter-layer fluctuations, which in turn are related to large fluctuations of the in-plane component of the orbital moment (see SM \cite{SI}): hence the large susceptibility ``jump'' at the IM transition. However, for a displacement field of \(V \approx \gamma_1/2\);
increasing $\Phi$ past $\Phi_c$ results in more localized states at the Fermi level, as can be seen in Fig.~\ref{fig:Bands} for $\Phi = 0.02$, reducing the orbital moment fluctuation and hence the weak paramagnetic response. Finally, one can see that if $V$ is not sufficiently large (check SM \cite{SI} for large displacement field: \(V>\gamma_1/2\)), that is to say, if the initial insulator at $\Phi=0$ is not sufficiently strong, the IM transition does not produce sufficient contrast between the two phases and the diamagnetic to paramagnetic crossover is eventually lost \cite{SI}.

Finally, we address the role of chemical potential in magnetic susceptibility by plotting $\chi$ as a function of $\mu$ for several values of $\Phi$ and $V$ in Fig.~\ref{fig:Chi_vs_mu}.  Additional gating and/or intentional doping can change the chemical potential. At zero magnetic field (panel (a)) the susceptibility is always negative (i.e., diamagnetic) at $\mu=0$, in agreement with Fig.~\ref{fig:Chi_vs_Phi}. As $\mu$ touches the band edge, $\chi$ changes sign and becomes paramagnetic as the Fermi contour contribution becomes dominant, due to the presence of delocalized states at the Fermi level. Further increase of $\mu$ reduces the magnitude of $\chi$, as the states at the Fermi level become more localized on one or the other layer.

At the critical magnetic field (Fig.\ref{fig:Chi_vs_mu}(b)), the zero-$\mu$ $\chi$ is strongly pushed in the positive direction, in agreement with Fig.~\ref{fig:Chi_vs_Phi}. This "push" results from the combined effect of delocalized states and the band edges, but it is not yet sufficient to create a paramagnetic response. The "push" diminishes as $\mu$ moves away from the charge neutrality point and states at the Fermi contour become more localized. Further increasing $\mu$ introduces another paramagnetic spike in $\chi$ when $\mu$ touches the edge of the higher conduction or lower valence bands.

Fig.\ref{fig:Chi_vs_mu}(c) shows the behavior of the susceptibility in the metallic state.  The remarkable feature is that paramagnetism now occurs even at the charge neutrality point.  We interpret this as the result of increasing delocalization of states at the Fermi level, which generates larger fluctuations of the in-plane moment (details in SM \cite{SI}). 
\begin{figure}
    \centering
    \includegraphics[width = \columnwidth]{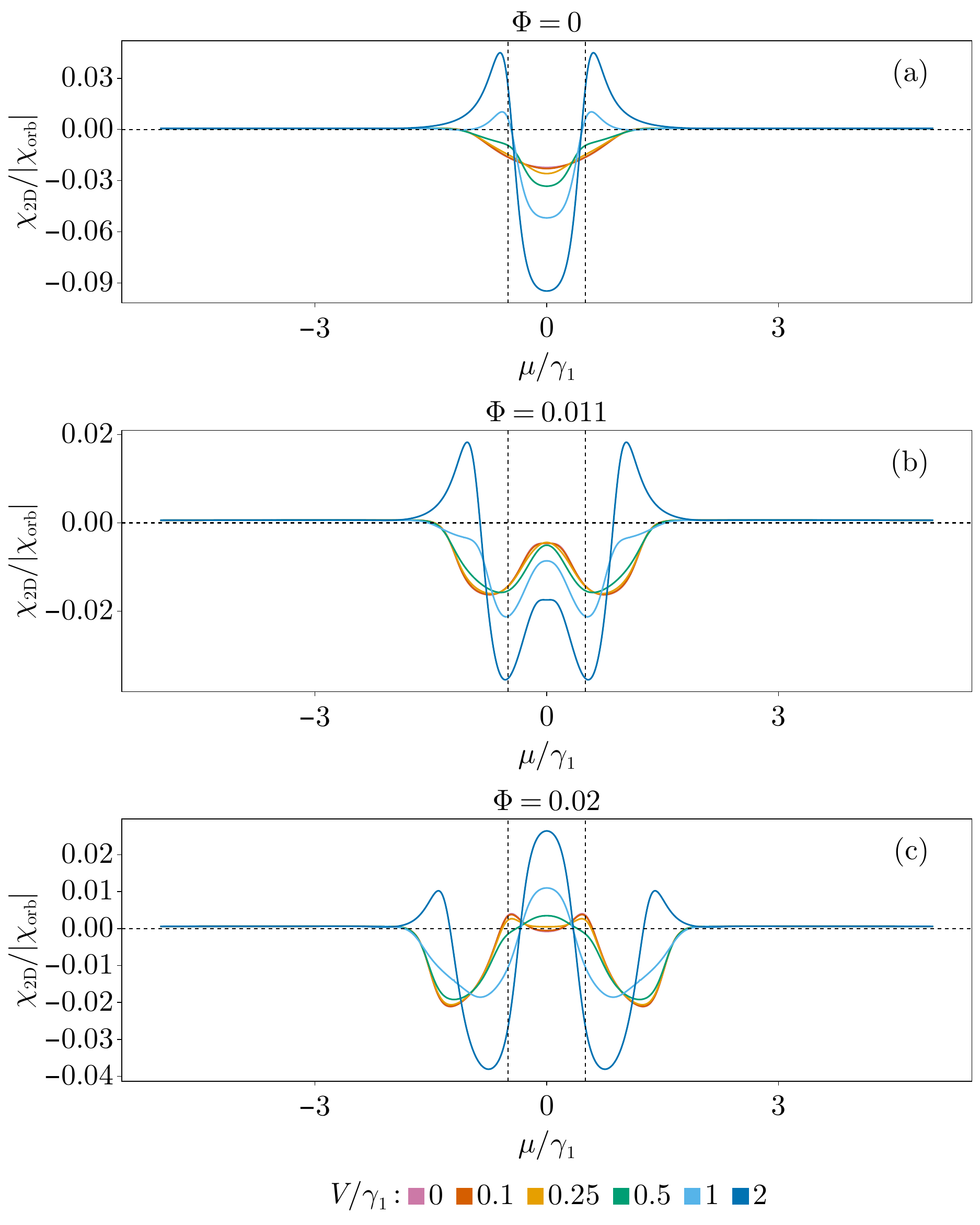}
    \caption{
    Orbital magnetic susceptibility $\chi$ (in units of $\alpha = \mu_0 (l / 2\pi \Phi_0)^2$) as a function of $\mu$ for different values of $V$ and $\Phi$ with $\theta = \pi / 2$.
    The calculations are performed at a finite temperature 
    \(T=300\)K .
}
    \label{fig:Chi_vs_mu}
\end{figure}

\textit{Observability.}
We have explored the possibility of altering the band structure, the transport, and the magnetic properties of stacked 2D materials by an in-plane magnetic field acting in conjunction with a vertical displacement field. While our calculations for bilayer graphene employ impractically high magnetic fields, we see considerable potential for reducing the required field through material engineering techniques. 

First of all, we notice that electron-electron interactions~\cite{RevModPhys.84.1067,doi:10.1126/science.aao2934} are predicted to enhance the slope of the single-layer cones by a potentially large numerical factor (logarithmically divergent in theory at the charge neutrality point).  While current experiments might be too disordered to observe this effect, it should be present in cleaner samples. This effect can considerably decrease the critical value of $\Phi$.

Second, applying a tensile P-strain pulling the layers apart \cite{doi:10.1021/jp300840k} can exponentially reduce the inter-layer coupling parameter. The strained inter-layer coupling strength is \(\gamma'_1= \gamma_1\*e^{-5.56*\epsilon_{p}}\), where \(\epsilon_{p}= (L-L_{0})/L_{0}\) quantifies the uniaxial strain,  \(L\) and \(L_0\) are, respectively, the deformed and undeformed inter-layer distances of the bilayer. According to this equation \(\gamma_1\) will be reduced by a factor $3$ when the strain is around \(20\%\),  and the critical field for the IM transition will be lowered to 270 T (without aid from the many-body effect mentioned above). 

Smaller critical fields can also be achieved by turning to artificial graphene structures \cite{Simon2012,Polini2013,PhysRevB.79.241406, Wang2018,wang2024} where the lattice constant is much larger and, accordingly, the band structure can be modified by relatively small magnetic fields.  Twisted multilayer graphene systems, such as twisted bilayer graphene~\cite{doi:10.1073/pnas.1108174108} and twisted double-bilayer graphene~\cite{PhysRevB.99.235417} are particularly promising candidates. In these systems, a mini (or Moiré) Brillouin zone emerges, and the distance between the nearest Dirac cones is reduced to approximately one-hundredth of its original value for small twist angles. This reduction allows effective manipulation of the band structure with magnetic fields on the order of 10 T. 
We note that the possibility of tuning the Dirac cones—bringing them together or pushing them apart—has already been experimentally demonstrated in twisted bilayer graphene~\cite{doi:10.1126/science.aaf4621}. 

In this  work we have  outlined a fundamental mechanism for the IM transition and magnetic switching in crossed fields within the simplest possible model. More complex systems, such as twisted multilayer graphene,  offer a promising direction for future research.

We thank K. S. Novoselov and for pointing us to the important  Ref.~\cite{doi:10.1126/science.aaf4621} and Hao Sun for useful discussions.
A.C. and G.V. were supported by the Ministry of Education, Singapore, under its Research Centre of Excellence award to the Institute for Functional Intelligent Materials (I-FIM, Project No. EDUNC-33-18-279-V12).

\bibliography{mybib}

\pagebreak
\setcounter{section}{0}
\setcounter{equation}{0}
\setcounter{figure}{0}
\setcounter{table}{0}
\newcounter{SIfig}
\setcounter{page}{1}
\makeatletter
\renewcommand{\theequation}{S\arabic{equation}}
\renewcommand{\thefigure}{S\arabic{figure}}
\renewcommand{\theSIfig}{S\arabic{SIfig}}
\renewcommand{\bibnumfmt}[1]{[S#1]}
\renewcommand{\citenumfont}[1]{S#1}

\begin{widetext}
\begin{center}
    \textbf{\large Supplemental Materials}\\
\textbf{ Insulator-Metal transition and magnetic crossover in bilayer graphene}
\end{center}
\end{widetext}






\maketitle
\title{Supplementary information for Insulator-Metal transition and magnetic crossover in bilayer graphene
}

\author{Amarnath Chakraborty}
\email{achakraborty@mail.missouri.edu}
\affiliation{Department of Physics and Astronomy, University of Missouri, Columbia, Missouri, USA}

\author{Aleksandr Rodin}
\email{aleksandr.rodin@yale-nus.edu.sg}
\affiliation{Yale-NUS College, 16 College Avenue West, 138527, Singapore}
\affiliation{Centre for Advanced 2D Materials, National University of Singapore, 117546, Singapore}
\affiliation{Department of Materials Science and Engineering, National University of Singapore, 9 Engineering Drive 1, 117575, Singapore}

\author{Shaffique Adam}
\affiliation{Department of Physics, Washington University in St. Louis, St. Louis, Missouri 63130, United States}
\affiliation{Department of Materials Science and Engineering, National University of Singapore, 9 Engineering Drive 1, 117575, Singapore}

\author{Giovanni Vignale}
\email{vgnl.g@nus.edu.sg}
\affiliation{The Institute for Functional Intelligent Materials (I-FIM), National University of Singapore, 4 Science Drive 2, Singapore 117544}
\date{\today} 

\maketitle
\section{Full-Tight Binding model of AB-Bilayer Graphene}
We present here the Bernal-stacked graphene bilayer with a parallel magnetic field and a transverse electric field described by a tight-binding Hamiltonian with nearest-neighbor intra- and interlayer hopping~\cite{PhysRevB.106.245143}
\begin{align}
H_{\mathbf{k}}& =\begin{pmatrix}
     V & \gamma_{0} f\left(\mathbf{k}_\mathrm{t}\right) & \gamma_4f\left(\mathbf{k}\right) & \gamma_3f^*\left(\mathbf{k}\right)\\
    \gamma_{0} f^{\dagger}\left(\mathbf{k}_\mathrm{t}\right) & V &\gamma_{1} &\gamma_4f\left(\mathbf{k}\right)\\
    \gamma_4f^*\left(\mathbf{k}\right) & \gamma_{1} & -V & \gamma_{0} f\left(\mathbf{k}_\mathrm{b}\right) \\
   \gamma_3f\left(\mathbf{k}\right) & \gamma_4f^*\left(\mathbf{k}\right) &\gamma_{0} f^{\dagger}\left(\mathbf{k}_\mathrm{b}\right) &-V
\end{pmatrix}
\label{eqn:FullHam}
\end{align}
In addition to the nearest-neighbor intralayer hopping energy $\gamma_0 = -2.8~\text{eV}$ and the vertical interlayer hopping $\gamma_1 = 0.3~\text{eV}$, we include the skew interlayer couplings $\gamma_3 = -0.1~\text{eV}$ and $\gamma_4 = 0.12~\text{eV}$ \cite{McCann_2013}. Using this model, we demonstrate that the insulator-to-metal (IM) transition predicted in the effective Hamiltonian discussed in the main text also occurs in the full tight-binding model. 

\begin{figure}[t]
    \centering
    \includegraphics[width = \columnwidth]{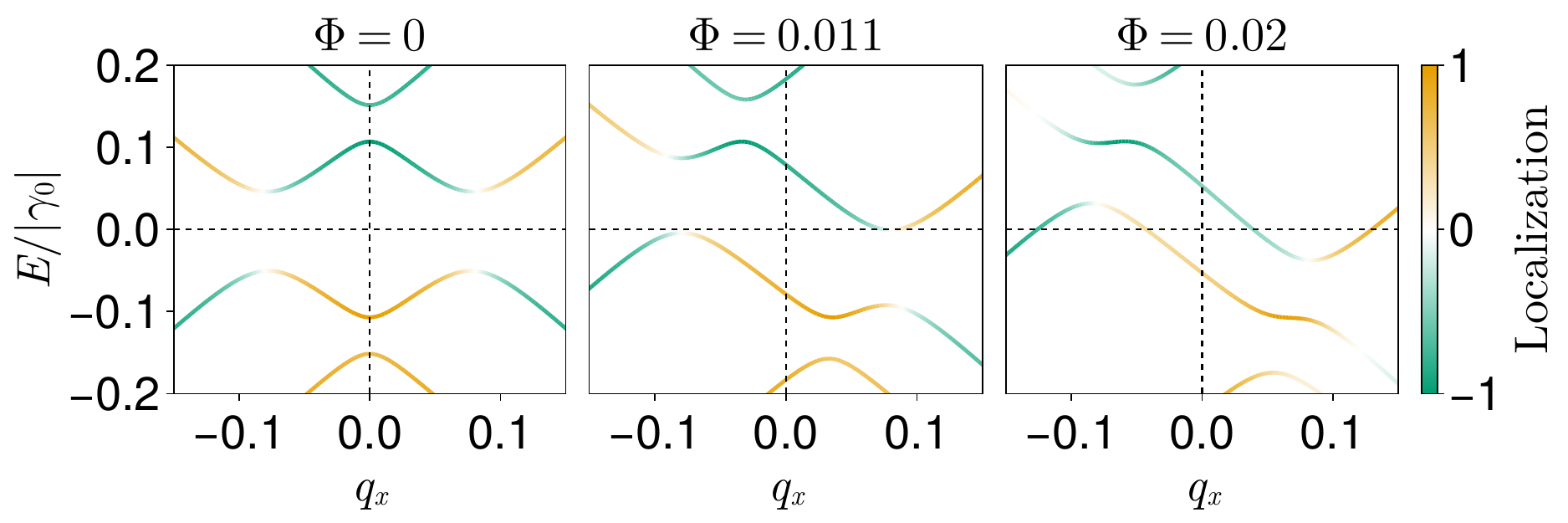}
    \caption{\textit{Full Tight-Binding Band edge.}We plot here the full-tight binding model including the skew-coupling terms \(\gamma_3,\gamma_4\). The three figures represent different in-plane magnetic fields and the corresponding localization of states for a displacement field of \(V=\gamma_1\). The central figure reveals that the IM transition happens at the same magnetic field even in the complete picture. We also kept the study of the localization built into the figures to show that the localization of states still exists at the band edge.}
    \label{fig:FullBand}
\end{figure}

In Fig.~\ref{fig:FullBand}, the central panel demonstrates that the IM transition occurs at the critical magnetic flux $\Phi_c = 0.011$, consistent with the results of the effective model. Notably, at $\Phi = \Phi_c$, the particle-hole symmetry is broken: the lower conduction band touches the Fermi level, whereas the upper valence band does not. Beyond the transition point, the pocket sizes of the valence and conduction bands become nearly identical.

This behavior is supported by the fact that the skew interlayer coupling terms are proportional to the factor $f_k$, which is unaffected by the applied magnetic field. \footnote{For interlayer hopping, the displacement vector between the two atoms lies predominantly in the out-of-plane direction \(\rv_{2j}-\rv_{1i} =d\mathbf{r} \to \hat{z} \). Since the vector potential of an in-plane magnetic field does not contribute in the out-of-plane direction, the line integral vanishes:
\(
\phi_{ij} = \frac{\hbar}{e} \int_{r_{1i}}^{r_{2j}} \mathbf{A} \cdot d\mathbf{r}\to 0.
\)
Thus, no effective phase shift arises for the interlayer coupling, leaving the hopping term \(f_k\) unchanged.} Furthermore, $f_k$ approaches zero near the Dirac points, resulting in a negligible contribution from these terms to the band structure.
 
In conclusion, the critical magnetic field for the IM transition remains numerically unchanged. Therefore, we continue to employ the simplified model described in the main text for further analysis without loss of generality.

\section{Model and gap-closing transition}
For the description of Bernal-stacked bilayer graphene in crossed magnetic and electric fields we employ  the well-known $4 \times 4$ tight-binding model
\begin{align}
    & H_\mathbf{k}(V,\Phi)=V\left(\tau_{z} \otimes \sigma_{0}\right)+\frac{\gamma_{1}}{2} \left(\tau_{+}\otimes\sigma_{-}+\tau_{-}\otimes\sigma_{+} \right)
    \nonumber
    \\
    & -\frac{\gamma_0}{2}\Big\{\tau_{0}\otimes\left[\fv\left(\mathbf{k}_\mathrm{t}\right)+\fv\left(\mathbf{k}_\mathrm{b}\right)\right] 
     +\tau_{z}\otimes\left[\fv\left(\mathbf{k}_\mathrm{t}\right)-\fv\left(\mathbf{k}_\mathrm{b}\right)\right] \Big\}\cdot\sigmav\,,
     \label{eqn:H_kproduct}
\end{align}
where the Pauli matrices $\boldsymbol{\sigma} = (\sigma_x, \sigma_y,\sigma_z)$, $\sigma_\pm = (\sigma_x \pm i \sigma_y) / 2$ act on the degree of freedom of the sublattice, the Pauli matrices  $\boldsymbol{\tau} = (\tau_x,\tau_y,\tau_z)$, $\tau_\pm = (\tau_x \pm i \tau_y) / 2$ act on the layer degree of freedom and $\otimes$ is the external product in the combined sublattice/layer space. Here $\mathbf{f}(\mathbf{k}) = (\mathrm{Re}\,f(\mathbf{k}), \mathrm{Im}\,f(\mathbf{k}),0)$. 
The in-plane magnetic field enters through the shifted wave vectors $\kv_t$ and $\kv_b$ defined for the $top$ and $bottom$-layer respectively. \(\mathbf{k}_\mathrm{t/b} a
    =  \mathbf{k} a \pm \pi\Phi (\sin\theta,-\cos\theta,0)\,,\)
   and $V$ is the electric displacement field. 
We note for future use the first and second derivative of the Hamiltonian with respect to the in-plane magnetic field:
\begin{align}
\begin{split}
    H'_{0,\mathbf{k}}(V,\Phi) &= -\frac{\gamma_0}{2}\Big\{\tau_{0}\otimes\left[\fv'\left(\mathbf{k}_\mathrm{t}\right)+\fv'\left(\mathbf{k}_\mathrm{b}\right)\right] \\
    & +\tau_{z}\otimes\left[\fv'\left(\mathbf{k}_\mathrm{t}\right)-\fv'\left(\mathbf{k}_\mathrm{b}\right)\right] \Big\}\cdot\sigmav\,,\\
 H''_{0,\mathbf{k}}(V,\Phi) &= -\frac{\gamma_0}{2}\Big\{\tau_{0}\otimes\left[\fv''\left(\mathbf{k}_\mathrm{t}\right)+\fv''\left(\mathbf{k}_\mathrm{b}\right)\right] \\
    & +\tau_{z}\otimes\left[\fv''\left(\mathbf{k}_\mathrm{t}\right)-\fv''\left(\mathbf{k}_\mathrm{b}\right)\right] \Big\}\cdot\sigmav\,,
\end{split}
\end{align}
where \(\fv'\) and \(\fv''\) represents the 1st and the 2nd derivative with respect to the in-plane magnetic field. At zero magnetic field, these expressions become much simpler 
\begin{align}
    \begin{split}
        H'_{\mathbf{k}}\left(V,\Phi=0\right) & \equiv \gamma_0 \tau_{z}\otimes \left[\fv' \cdot\sigmav \right]  ,\\
 H''_{\mathbf{k}}\left(V,\Phi=0\right) & \equiv \gamma_0 \tau_{0}\otimes\left[\fv''\cdot\sigmav \right]\,;
    \end{split}
\end{align}
It is easy to verify that for any $V$ and $\Phi$, the Hamiltonian changes sign under the unitary transformation $U_1 = \tau_x \otimes \sigma_y$  i.e., 
\begin{equation}\label{U1}
U_1 H_\mathbf{k}(V,\Phi) U^{\dagger}_1 = -H_{-\mathbf{k}}(V,\Phi)\,.   
\end{equation} 
In addition, the unitary transformation $U_2=\tau_0 \otimes \sigma_z$
reverses the sign of the Hamiltonian {\it and} the electric potential:  
\begin{equation} \label{U2}
U_2 H_\kv(V,\Phi) U_2^{\dagger} = -H_\kv(-V,\Phi)\,.
\end{equation}
Finally, $H_\kv(V,\Phi)$ and $H_{-\kv}(V,-\Phi)$ have the same spectrum: in fact, these two hermitian operators are the complex conjugate of each other.  
Knowing that  \(k\to -k\)  changes the sign of the spectrum (Eq.~\eqref{U1})  we can safely conclude that \(\Phi \to -\Phi\)  (at fixed $\kv$) also changes the sign of the spectrum. These symmetries imply that the gap is centered around zero energy and closes when a valence and a conduction band become degenerate at zero energy, leading to the insulator-to-metal (IM) transition.
%
The existence of zero-energy states can be assessed by equating the determinant of Eq.~\eqref{eqn:H_kproduct} to zero:
\begin{align}
    |H_\mathbf{k}(V,\Phi)| &= V^4 + \left[\gamma^2_{1}-\gamma^2_{0}\left(\left|f\left(\mathbf{k}_\mathrm{t}\right)\right|^2+\left|f\left(\mathbf{k}_\mathrm{b}\right)\right|^2\right)\right]V^{2}  
    \nonumber
    \\ &+\gamma^4_{0}\left|f\left(\mathbf{k}_\mathrm{t}\right)\right|^2 \left|f\left(\mathbf{k}_\mathrm{b}\right)\right|^2 = 0\,.
    \label{eqn:seceqn}
\end{align}
Solving Eq.~\eqref{eqn:seceqn} for $V^2$ yields two solutions
\begin{align}
    V_\pm^2 &= \frac{1}{2}\left(\frac{C_+ + C_-}{2}\pm\sqrt{C_+C_-}\right)\,,
    \label{eqn:V_solution}
    \\
    C_\pm &= \gamma_0^2\left[|f(\mathbf{k}_\mathrm{t})|\pm |f(\mathbf{k}_\mathrm{b})|\right]^2 - \gamma_1^2\,.
    \label{eqn:C}
\end{align}
The existence of real roots $V_\pm$ for a certain value of $\Phi$ at any point within the Brillouin zone indicates that there are zero-energy states and the gap is closed. As can be seen from Eq.~\eqref{eqn:V_solution}, the necessary and sufficient condition for the existence of real $V_\pm$ is $C_+ > C_-\geq 0$.  By determining the minimum value of $\Phi$ for which a region with $C_->0$ appears in the Brillouin zone, we determine the minimum magnetic field that can close the gap opened by the transverse electric field. We observe that the band closure takes place close in the vicinity of the K point, which allows us to expand $C_-$ around it and equate $C_-$ to zero, leading to
\begin{equation}
    \frac{9}{4}\left[|\mathbf{k}_\mathrm{t}| - |\mathbf{k}_\mathrm{b}|\right]^2= \frac{\gamma_1^2}{\gamma_0^2}\,.
    \label{eqn:Contour_equlaity}
\end{equation}
The maximum value the quantity inside the brackets can take is $\pi \Phi / \Phi_0$, leading to $\Phi = |\gamma_1 / 3\pi\gamma_0|\Phi_0 \approx 0.011 \Phi_0$ as the minimum value of $\Phi$ required to close the gap.

\section{Localization of states and Magnetic response}
\begin{figure}
    \centering
    \includegraphics[width = \columnwidth]{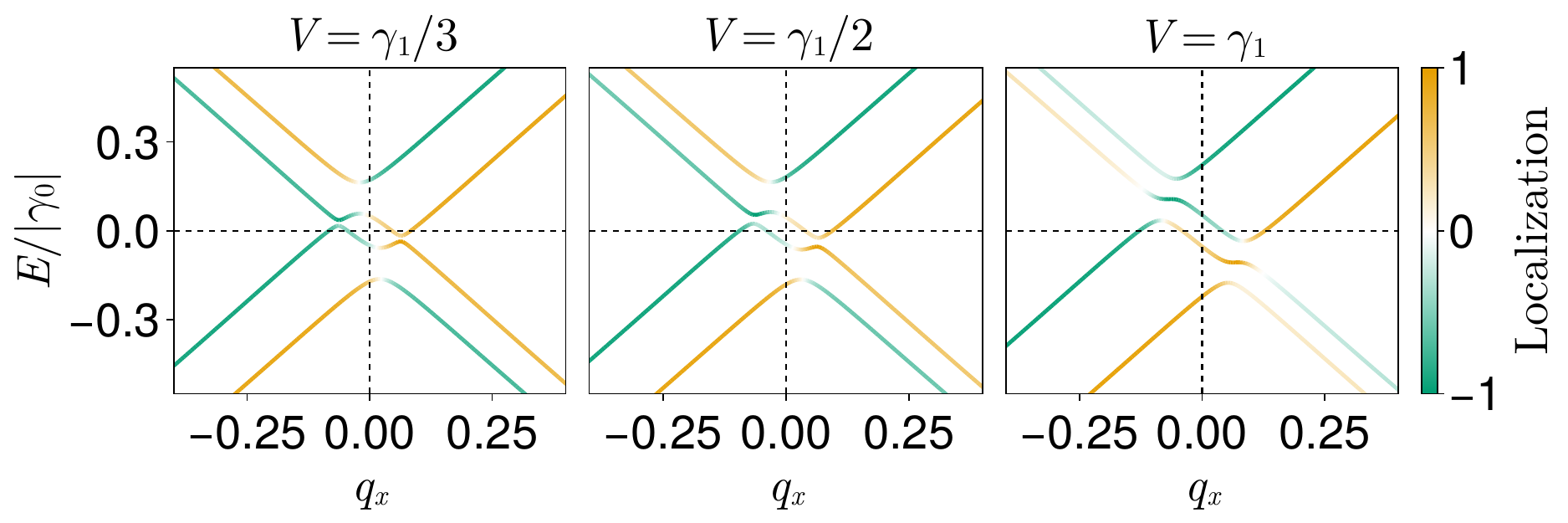}
    \caption{\textit{Band edge and Localization.} The three figures represent different displacement fields and the corresponding localization of states for an in-plane magnetic field of \(\Phi=0.02\) (\(\Phi \geq \Phi_c\)). Our localization study reveals that for \( V \geq \gamma_1 \), the band minima (lower conduction band) consist of delocalized particles, which, along with the band edge contribution, are responsible for the paramagnetic response at \(\mu=0\). However, for \( V \leq \gamma_1/2 \), the deep states and localized states no longer coincide. Consequently, the paramagnetic response is significantly weakened.}
    \label{fig:Band_loc2}
\end{figure}

\textit{Localization}: To justify our results of the magnetic response we will take into account two factors; one is the localization of the states and the other is the displacement field. Let us first define what we mean by localization of states:
\begin{equation}
    L(k, \Phi, V)= \left[\left(|\psi_{1A}|^2+|\psi_{1B}|^2\right)-\left(|\psi_{2A}|^2+|\psi_{2B}|^2\right)\right]
\label{eqn:Local}
\end{equation}
here the \(\phi\)'s are the wavefunction of their respective layer and sublattices. For delocalized states, the localization function approaches zero. To understand band localization, we examined the region near the Dirac cones and found delocalized states near the band minima. Particles in these states, when crossing the chemical potential thresholds, become susceptible to magnetic fields.

Carefully looking at the Eq.~\eqref{eqn:Local}; we can state that the Localization function is the expectation value of the operator \(\tau_z\). 
\begin{equation}
    L = \langle \tau_z \rangle
\end{equation}
and the mean squared fluctuation of z is 
\begin{equation}
    \left(\Delta\tau_Z\right)^2= \langle \tau_z^2 \rangle -\langle \tau_z\rangle ^2= 1-L^2.
\label{eqn:Fluctuation in z}
\end{equation}
thus for the delocalized states (\(L\approx 0\)) the mean-squared fluctuations in \(\tau_z\) is close to 1. Hence the fluctuation is highest for delocalized states, and any slight change in the localization factor will significantly affect the fluctuation. 

\textit{Magnetic Moment}: Next, we will examine the in-plane orbital moment and correlate it with the localization of states. This analysis will strengthen our argument that the change in localization, as we move around the band, is responsible for the observed changes in the system's magnetic response.

The equation that describes the in-plane magnetic moment is as follows
\begin{equation}
    \mv= \frac{e}{4\hbar} \left[\rv \cross \vv -\vv \cross \rv \right]; 
\end{equation}
here \(\vv\) refers to the velocity operator with respect to the applied field. 
If we consider the magnetic field is along the \(y-\)direction, then the moment can be written as 
\begin{equation}
    m_y = \frac{e}{2\hbar}  z v_x 
\end{equation}
the \(x v_z\) term contributes the same as \(z v_x\) , but with a negative sign.
Then, using 
\begin{equation}
    z=\tau_z: ~~~v_x=\partial H/\partial k_x = \gamma_0  \tau_0 (\partial \fv /\partial k_x).\sigmav,  
\end{equation}
this above equation is valid for zero field calculations only; otherwise, for non-zero field moment calculations, the Fermi velocities will be different in each layer. In the zero-field case, we have:
\begin{equation}
    m_y = \gamma_0 \tau_z \left(\partial \mathbf{f} / \partial k_x \cdot \boldsymbol{\sigma}\right) = H'_{0,\mathbf{k}}
\label{eqn:moment}
\end{equation}
From this equation, we observe that the in-plane moment in the zero-field case is directly proportional to the final term of the susceptibility, namely the Fermi surface term (responsible for the paramagnetic response at charge neutrality). Additionally, the moment is proportional to the \(\tau_z\)-operator, implying that a high fluctuation in the \(\tau_z\)-operator will result in a high fluctuation in the in-plane magnetic moment. 
 
Our observations indicate that the fluctuation in the \(\tau_z\)-operator is highest for delocalized states. Therefore, we conclude that delocalized states induce the greatest fluctuation in the in-plane moment, leading to a stronger paramagnetic response. \footnote{Classically the paramagnetic response is proportional to the mean squared fluctuation of the magnetic moment; \(\chi = n\left(\Delta m_y\right)^2 /T\) (n is the electron density, T the temperature). This gives the correct quantum result for the electron gas in the \(T\to 0\) limit.} 

Having established that high fluctuations in the localization of states are responsible for the paramagnetic response in susceptibility, we now discuss the emerging scenarios and their corresponding explanations. The radius of these pockets is influenced by the displacement field, which, when increased, encompasses more delocalized states. In our discussion, we refer to these sufficiently dense states as deep states. When deep states coincide with delocalized states, a paramagnetic response emerges.

The deep states are located at the local extremas of the lower conduction (higher valence) band, which are approximately at \( E \approx \frac{\gamma_1}{2} \) or \( E \approx 0 \). If these locations have delocalized states (\( L \approx 0 \)), it results in high fluctuations in the magnetic moment, leading to a paramagnetic response. Assuming this to be true, let us analyze the magnetic response observed in Fig.3 of the main text and the band dispersion observed at Fig.~\ref{fig:Band_loc2}.

\begin{itemize}
    \item For \( V = \frac{\gamma_1}{3} \) at \( E \approx 0 \), there are delocalized states that are not deep enough for \( \Phi < \Phi_c \). However, for \( \Phi > \Phi_c \) at \( E \approx \frac{\gamma_1}{2} \), the coincidence of delocalized states with the band edge is observed, but they are not deep enough, leading to a weak response.
    
    \item For \( V = \frac{\gamma_1}{2} \) at \( E \approx 0 \), the delocalized states are not deep enough for \( \Phi < \Phi_c \) resulting a response that is reaching towards the paramagnetism. For \( \Phi > \Phi_c \), even though the states are deep enough, they are not sufficiently delocalized, resulting in a weak paramagnetic response.
    
    \item For \( V \geq \gamma_1 \), there are highly dense delocalized states at \( E \approx 0 \) for \( \Phi > \Phi_c \). For \( \Phi < \Phi_c \), the delocalized states are not deep enough.
    
\end{itemize}

The key point is that a significant paramagnetic response occurs when there is a deep enough band edge (\( V \geq \gamma_1 \)) along with delocalized states (\( L \approx 0 \)). If either of these criteria is not adequately met, the paramagnetic response will be weak or absent.

\end{document}